\documentclass[11pt]{article}
\usepackage[margin=1in]{geometry}
\usepackage{graphicx}
\usepackage{amsmath}
\usepackage{xcolor}
\usepackage{mathtools}
\usepackage{tikz}
\usepackage{float}
\usepackage{hyperref}

\usetikzlibrary{positioning}
\usetikzlibrary{circuits.logic.US,circuits.ee.IEC,positioning,fit,arrows,decorations.pathmorphing,backgrounds}

\def\beq{\begin{equation}}
\def\eeq#1{\label{#1}\end{equation}}
\def\eeqn{\end{equation}}

\def\beqa{\begin{eqnarray}}
\def\eeqa#1{\label{#1}\end{eqnarray}}
\def\eeqan{\end{eqnarray}}

\let\bar=\overbar

\def\Dslash{\not{\hbox{\kern-4pt $D$}}}
\def\dslash{\not{\hbox{\kern-2pt $\del$}}}

\def\msb{{\bar{\ssstyle M \kern -1pt S}}}

\def\Title#1{\begin{center} {\Large {\bf #1} } \end{center}}
\def\Author#1{\begin{center} {\normalsize {\sc #1} } \end{center}}
\def\Institution#1{\begin{center} {\normalsize {\it #1} } \end{center}}
\def\Abstract#1{\noindent {\normalsize {\bf Abstract:} {\normalfont #1}}}
\def\Conference{\vspace{4mm}\begin{raggedright} {\normalsize {\it Talk presented at the 2019 Meeting of the Division of Particles and Fields of the American Physical Society (DPF2019), July 29--August 2, 2019, Northeastern University, Boston, C1907293.} } \end{raggedright}\vspace{4mm}}

\begin{document}

%
%

\Title{Development and Characterization of Solid Noble Bolometers}

\Author{Philip L. R. Weigel\footnote{Corresponding author: plw38@drexel.edu}, Erin V. Hansen, Michelle J. Dolinski}

\Institution{Department of Physics\\ Drexel University, Philadelphia, PA, USA}

\Abstract{Noble liquid detectors have become an attractive option for exploring physics beyond the standard model. Current experiments are using these detectors to search for dark matter interactions, neutrinoless double beta decay, and other phenomena. Improved energy resolution can be leveraged from an optimized combination of two detection channels: ionization and scintillation. Experimentally, a microscopic anti-correlation behavior between these signals has been observed, but it has not been described from first principles. Making measurements in a third channel would provide useful information about the microscopic anti-correlation phenomenon. Work is currently underway at Drexel University to develop solid argon and xenon bolometers, which would be able to utilize a heat channel in addition to ionization and scintillation. Present efforts are aimed at developing a method for growing small noble solid samples via vapor deposition onto a substrate over a wide range of temperatures down to 10 K. Understanding the sample growth is the first step to develop techniques for integrating detector components to measure ionization and scintillation signals. This will allow for improved characterization of noble solids as detector media. In the future, efforts will be focused on the growth of these detectors in the Drexel dilution refrigerator, where samples can be cooled to 20 mK to include bolometric measurements for the simultaneous readout of the three detection channels.}

\Conference

\section{Introduction}

In the past few decades, cryogenic noble liquid detectors have become a popular technology in the search for rare events, such as dark matter interactions \cite{Akerib:2015cja, Aprile:2017aty, Agnes:2018fwg}, neutrinoless double beta decay \cite{Albert:2017owj, Kharusi:2018eqi}, and neutrino interactions \cite{Acciarri:2016smi, Acciarri:2015uup}. When a particle interacts within the medium of a noble liquid-type detector, energy is eventually transferred into three channels: light, charge, and heat. Photons are produced by the decay of eximers, and by the recombination of an ion with a free electron. Xenon scintillates at $\sim$175 nm, and argon at $\sim$130 nm. Liberated electrons in a noble liquid are free to drift until they are captured by electronegative impurities. The third channel, heat, cannot be measured since the thermal energy is small compared to thermal fluctuations at the operating temperature. Many of the current experiments utilize both scintillation and ionization channels.

A microscopic anti-correlation effect has been observed in noble liquid detectors, most notably in liquid xenon, between light and charge yields \cite{Conti:2003av, Aprile:2007qd}. This is caused by correlated fluctuations in the partitioning of energy into charge and light for each event, but the sum is constant. This effect can be used to provide improved overall energy resolution compared to the resolution of individual channels by creating a rotated energy axis in the light-charge plane. For an event with collected light energy $E_{s}$ and charge energy $E_{q}$, the combined energy can be expressed as

\begin{equation}
    E_{c} = \frac{E_{s}\sin\theta + E_{q}\cos\theta}{\sin\theta + \cos\theta}
\end{equation}

\noindent where $\theta$ is the chosen rotation angle for the rotated energy axis \cite{Aprile:2007qd}. The power of performing this interaction lies in the significant improvement of the combined energy resolution. For example, EXO-200 obtains an overall energy resolution of 1.15\% with resolutions of 4.80\% and 2.84\% for the light ($R_{s}$) and charge ($R_{q}$) channels respectively \cite{EXO_res}. The resolution with a rotation angle $\theta$ is determined by

\begin{equation}
R_{c}^2 = \frac{R_{s}^{2} \sin^2 \theta + R_{q}^2 \cos^2 \theta + 2 R_{sq} \sin \theta \cos \theta }{(\sin \theta + \cos \theta)^2}
\end{equation}

\noindent where $R_{sq}$ is related to the correlation coefficient of the scintillation and ionization channels $\rho_{sq}$ by

\begin{equation}
    R_{sq} = \rho_{sq} R_{s} R_{q}
\end{equation}

In the liquid phase, all energy released into the heat channel is lost. A measurement of all three energy channels would allow for the study of the correlated fluctuations in more detail, with potential applications to current noble liquid detectors. This can be done by taking xenon down to very low temperatures ($\sim$tens of mK), where temperature variations are measureable. In this regime, it should be possible to measure light, charge, and phonon yields simultaneously. The technology for operating cryogenic bolometers for the purpose of rare event searches has been around for decades \cite{Shutt:1992qg, Pessina:1992ec}. These experiments typically use single crystals cooled to tens of mK and equipped with phonon readout devices. However, it is technically challenging to grow a single crystal of xenon or argon \cite{Baoto}. Recent experiments at Fermilab were successful in growing a kilogram-scale optically transparent solid xenon sample at 157 K \cite{Yoo:2015xza}, but no research has been done to evaluate this medium at the temperatures required for a phonon readout. Systematic studies of a solid argon and xenon bolometers would provide useful information about the detector microphysics in these media. 

At Drexel University, we are current developing a test stand to grow solid xenon samples on the scale of a few grams ($\sim$1-2 cm$^3$). These tests will initially be carried out over a range of temperatures controlled by closed-cycle cryocooler capable of reaching temperatures down to 4.2 K. This stage has two primary goals: to find the optimal procedure for growing samples, and to take measurements of light and charge yields in noble solids. The first samples are expected by late 2019. The work described here is focused on the thermal modeling of a solid xenon bolometer and on determining the thermal conductance at the boundary between xenon and other materials. This information is needed to design an optimal thermal sensor and readout for a noble solid bolometer.

\section{Bolometer Design}

Instrumenting the phonon channel of a noble solid bolometer could provide useful information about energy partitioning in noble liquid detectors. To do this, a noble solid must be cooled to sub-Kelvin temperatures in order to obtain a measurable signal from keV or MeV-scale energy depositions. There are currently a wide range of temperature sensors utilized as a readout for macrobolometers, such as transition edge sensors (TES) or microwave kinetic inductance detectors (MKID). For the application of a solid xenon bolometer, we have decided that germanium neutron transmutation doped (NTD) thermistors will be used. The reason for this choice is that for polycrystalline samples, the fast athermal phonon signal is lost, but the slower thermalized phonon signal can be easily read out. NTD thermistors have been well studied for the application of cryogenic macrobolometers \cite{Pedretti, Wang:1989vk, Vignati}.

For a typical readout of the thermistor, a bias voltage $V_{b}$ is applied through a load resistor $R_{L}$, which in turn is in series with the parallel combination of the thermistor $R(T)$ with a parasitic capacitance $c_{p}$. For a germanium NTD thermistor, the resistance as a function of electron temperature ($T_{e}$) is given by

\begin{equation}
    \label{rte}
    R(T_{e}) = R_{0} e^{(\tfrac{T_0}{T_{e}})^{\gamma}}
\end{equation}

Parameters $R_{0}$, $T_{0}$ and $\gamma$ (approximately equal to $1/2$) are characteristic parameters of the thermistor. As the thermistor resistance varies with temperature, the voltage across it will vary. This voltage can then be sent through an amplifier, filtered, and digitized to obtain a readout of thermal signals in the bolometer.

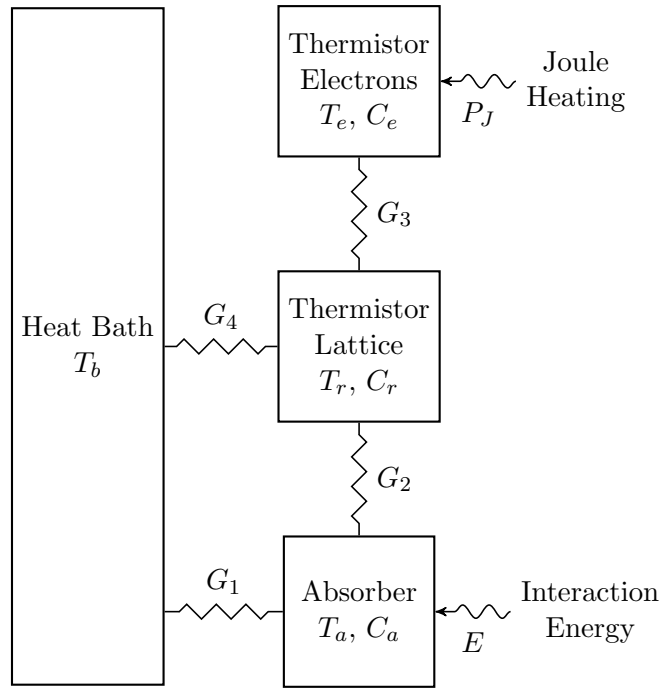
\begin{figure}[H]

    \centering
    \begin{tikzpicture}[
    bolnode/.style={rectangle, draw=black, fill=white, thick, minimum size=2cm},
    bathnode/.style={rectangle, draw=black, fill=white, thick, minimum width=2 cm, minimum height=9cm},
    circuit ee IEC,circuit logic US,x=3cm,y=2cm,semithick,
    set resistor graphic=var resistor IEC graphic,
    >=stealth',
    photon/.style={decorate,decoration={snake,post length=1mm}}, 
    ]
    
        \node[bolnode, align=center] (absorber) {Absorber\\$T_{a}$, $C_{a}$};
        \node[bolnode, align=center] (lattice) [above=1.5 cm of absorber] {Thermistor\\Lattice\\$T_{r}$, $C_{r}$};
        \node[bolnode, align=center] (elec) [above=1.5cm of lattice] {Thermistor\\Electrons\\$T_{e}$, $C_{e}$};
        
        \node[bathnode, align=center] (bath) [left=1.5cm of lattice] {Heat Bath\\$T_{b}$};
        
        \node[right=1 cm of absorber, align=center] (e_in) {Interaction\\Energy};
        \node[right=1 cm of elec, align=center] (joule) {Joule\\Heating};
        
        \draw (absorber) to [resistor={info'=$G_{2}$}] (lattice);
        \draw (absorber) to [resistor={info'=$G_{1}$}] (absorber-|bath.east);
        \draw (lattice) to [resistor={info'=$G_{3}$}] (elec);
        \draw (lattice) to [resistor={info'=$G_{4}$}] (bath);
        
        \draw[->,photon] (e_in)  -- node[label=below:$E$] {} (absorber);
        \draw[->,photon] (joule) -- node[label=below:$P_{J}$] {} (elec);
    
    \end{tikzpicture}
    
    \caption{Bolometer schematic. The energy absorber has a heat capacity $C_{a}$ with temperature $T_{a}$ and is thermally linked to a heat bath with temperature $T_{b}$ with a thermal conductance $G_{1}$. In addition to the link to the heat bath, the absorber is also in thermal contact with the atomic lattice of the NTD thermistor. This conductance is represented by $G_{2}$. The temperature of the thermistor atomic lattice is represented by $T_{r}$, and it has a heat capacity $C_{r}$. Then, the thermistor lattice is thermally tied to the heat bath (by wires or other means) by $G_{4}$ and to the thermistor electrons by $G_{3}$. The electronic heat capacity is given by $C_{e}$.}
    \label{bolometer_figure}
\end{figure}

The bolometer model used here (Figure \ref{bolometer_figure}) follows that in \cite{Pedretti}, and for the rest of this section, the absorber material will be assumed to be solid xenon. Heat capacities for each element are denoted by $C$, thermal conductivities by $G$, and temperatures $T$. It is convenient to put the conductivities, which typically depend on temperature to some power $\alpha$, in the following form:

\begin{equation}
    G_{i} (T) = g_{i} T^{\alpha_{i}}
\end{equation}

With no bias voltage, the temperatures of all the bolometer components lower to the bath temperature. By introducing the heat contribution from Joule heating of the thermistor, this is not the case. The power dissipated by the thermistor depends on the thermistor electron temperature (as R depends on it), and applied voltage $V$:

\begin{equation}
    P_{J} = R(T_{e}) \Big[\frac{V}{R(T_{e}) + R_{L}} \Big]^2
\end{equation}

Balancing the transfer of thermal energy in the system, we arrive at a set of equations that describe the changes in temperatures of each component. Additional powers that account for noise can be introduced ($P_{b,e}$ and $P_{a}$).

\begin{align}
    C_{e}\dot{T_{e}} & = P_J + P_{b,e} - \frac{g_3}{\alpha_3 + 1}(T_{e}^{\alpha_3 + 1} - T_{r}^{\alpha_3 + 1}) \\
    C_{r}\dot{T_{r}} & = 
    \frac{g_3}{\alpha_3 + 1}(T_{e}^{\alpha_3 + 1} - T_{r}^{\alpha_3 + 1}) + 
    \frac{g_2}{\alpha_2 + 1}(T_{a}^{\alpha_2 + 1} - T_{r}^{\alpha_2 + 1}) - 
    \frac{g_4}{\alpha_4 + 1}(T_{r}^{\alpha_4 + 1} - T_{b}^{\alpha_4 + 1})\\
    C_{a}\dot{T_{a}} & = P_{a} - \frac{g_2}{\alpha_2 + 1}(T_{a}^{\alpha_2 + 1} - T_{r}^{\alpha_2 + 1})
    - \frac{g_1}{\alpha_1 + 1}(T_{a}^{\alpha_1 + 1} - T_{b}^{\alpha_1 + 1})
\end{align}

In addition to these three equations, there is another that accounts for the dynamic behavior of the electrothermal feedback of the thermistor.

\begin{equation}
    \dot{V} = \frac{1}{c_{p} R_{L}} \Big[ V_{bias} - \frac{V(R(T_{e}) + R_{L})}{R(T_{e})} \Big]
\end{equation}

This system of coupled differential equations can be solved with many different numerical integration routines. The initial conditions are found by setting all the temperatures to the bath temperature and solving the system with no interaction energy. It should be noted that this method of calculation for the thermal and electrical response of the bolometer system is only being used as an order of magnitude estimate for the feasibility of noble solids as the absorber. It is desirable to show that the decay time of the thermistor signal is fast with respect to expected event rates.

\begin{figure}[t]
    \centering
    \includegraphics[width=0.75\textwidth]{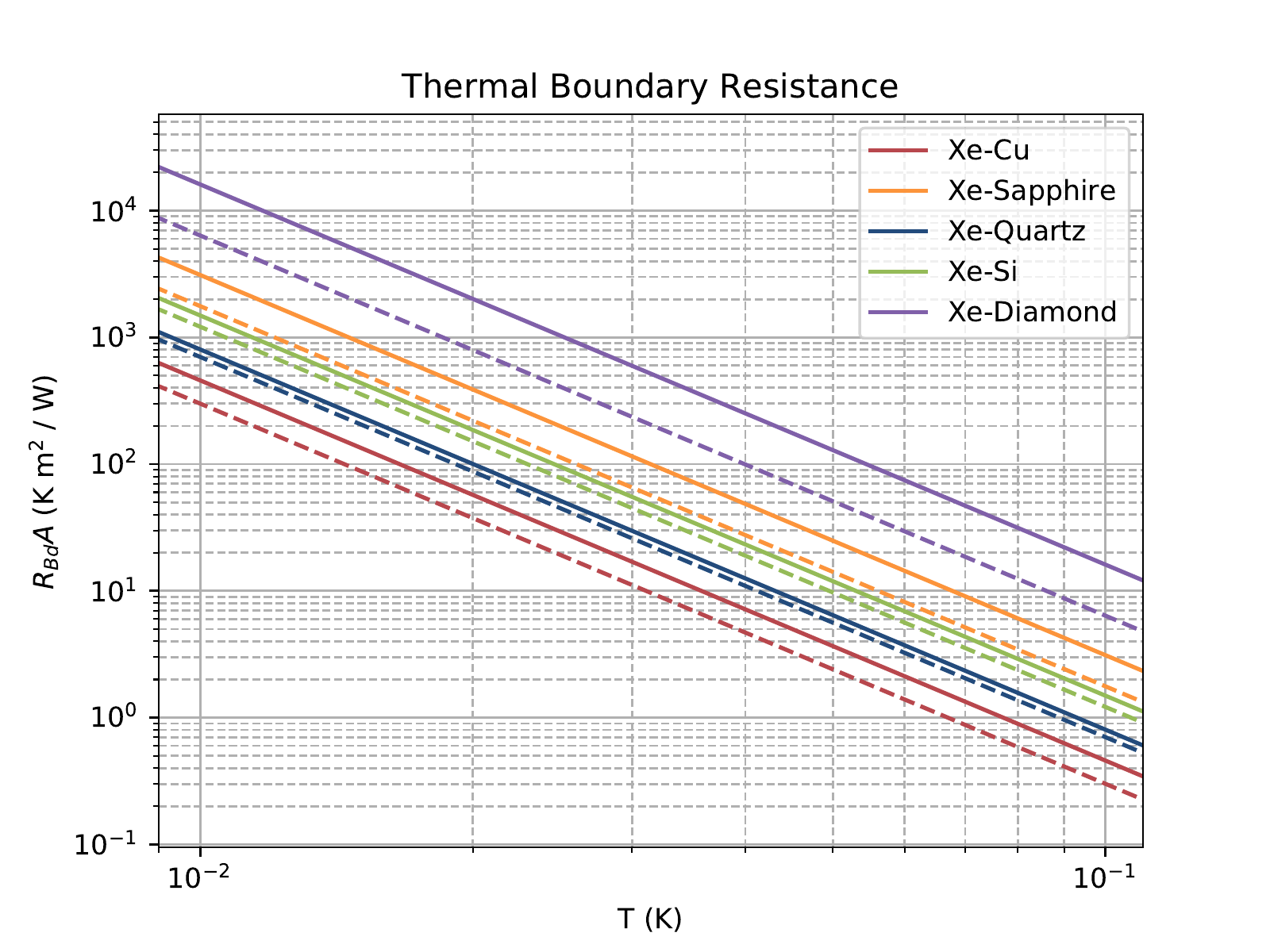}
    \caption{Calculated thermal boundary resistance between xenon and other materials. Solid lines are calculations with AMM, and dashed lines are calculations with DMM.}
    \label{tbr}
\end{figure}

The xenon sample will be grown on a substrate, and so it will be in relatively good thermal contact with that surface. The substrate is assumed to be at the heat bath temperature. Little data exists for the conductances between argon/xenon and other materials, but models exist to calculate it. The two models that will be considered here are the acoustic mismatch model (AMM) and diffuse mismatch model (DMM). In the AMM, phonons are treated as acoustic waves that can only reflect, refract, or mode convert. For this model, the thermal boundary resistance depends on the the speeds of sound $v_{i, j}$ for material $i$ and mode $j$, and also transmission probabilities $\Gamma$. DMM assumes that all phonons lose all information about where they came from as the go from material 1 to material 2 and depends only on the speeds of sound. These calculations are discussed in significant detail in \cite{Swartz:1989zz}.

\begin{table}
    \begin{center}
        \begin{tabular}{l|ccc} \hline 
        Material & $\rho$ (g/cm$^3$) &  v$_{l}$ (m/s) &  v$_{t}$ (m/s) \\ \hline
         Copper & 8.96 & 4910 & 2500 \\
         Diamond & 3.512 & 17500 & 12800 \\
         Quartz & 2.66 & 6090 & 4100 \\
         Sapphire & 3.97 & 10890 & 6450 \\
         Silicon & 2.33 & 8970 & 5332 \\
         Xenon & 3.781 & 1150 & 873 \\ \hline
        \end{tabular}
        \caption{Values of the density $\rho$, longitudinal speed of sound v$_{l}$, and transverse speed of sound v$_{t}$ for each of the materials used in Figure \ref{tbr}. All values except for xenon are from \cite{Swartz:1989zz}. The value for xenon density is from \cite{Sears} and speeds of sound are from \cite{Lurie}.}
        \label{tab:physical_vals}
    \end{center}
\end{table}

\begin{equation}
    \label{AMM_DMM}
    G_{Bd}^{\textrm{AMM}} = \frac{\pi^2 k_{b}^{4}}{15\hbar^3} \sum_{j} \frac{\Gamma_{1,j} }{v_{1,j}^2} T^3, \quad \quad
    G_{Bd}^{\textrm{DMM}} = \frac{\pi^2 k_{b}^{4}}{30\hbar^3} \frac{(\sum_j v_{1,j}^{-2})(\sum_j v_{2,j}^{-2})}{\sum_{i, j} v_{i,j}^{-2}} T^3
\end{equation}

It is recommended by \cite{Swartz:1989zz} that AMM and DMM should be used upper and lower limits of the true thermal boundary conductance for solid-solid interfaces below 30 K. Other models have been developed to include more effects such as the true phonon density of states and dispersion relations \cite{Reddy}, but these are not included in this study. Since the Debye phonon density of states was used for these calculations, both AMM and DMM have a cubic dependence on temperature ($\alpha = 3$). 
The conductance between xenon and potential substrate materials were calculated for both AMM and DMM using the values for density and speeds of sound from Table \ref{tab:physical_vals}. The results from xenon are shown in Figure \ref{tbr}, and the same process can be repeated with the physical constants of argon. Beyond the capabilities of the models used here, there is considerable interest in using polymers and sintered ceramics (i.e. PTFE and alumina) as weak thermal links \cite{Drobizhev:2016qbx}. PTFE is desirable for its good reflectivity of 175 nm light from xenon scintillation \cite{silva}, but suffers from mechanical issues related to thermal contraction at low temperatures. 

\begin{figure}
    \centering
    \includegraphics[width=0.75\textwidth]{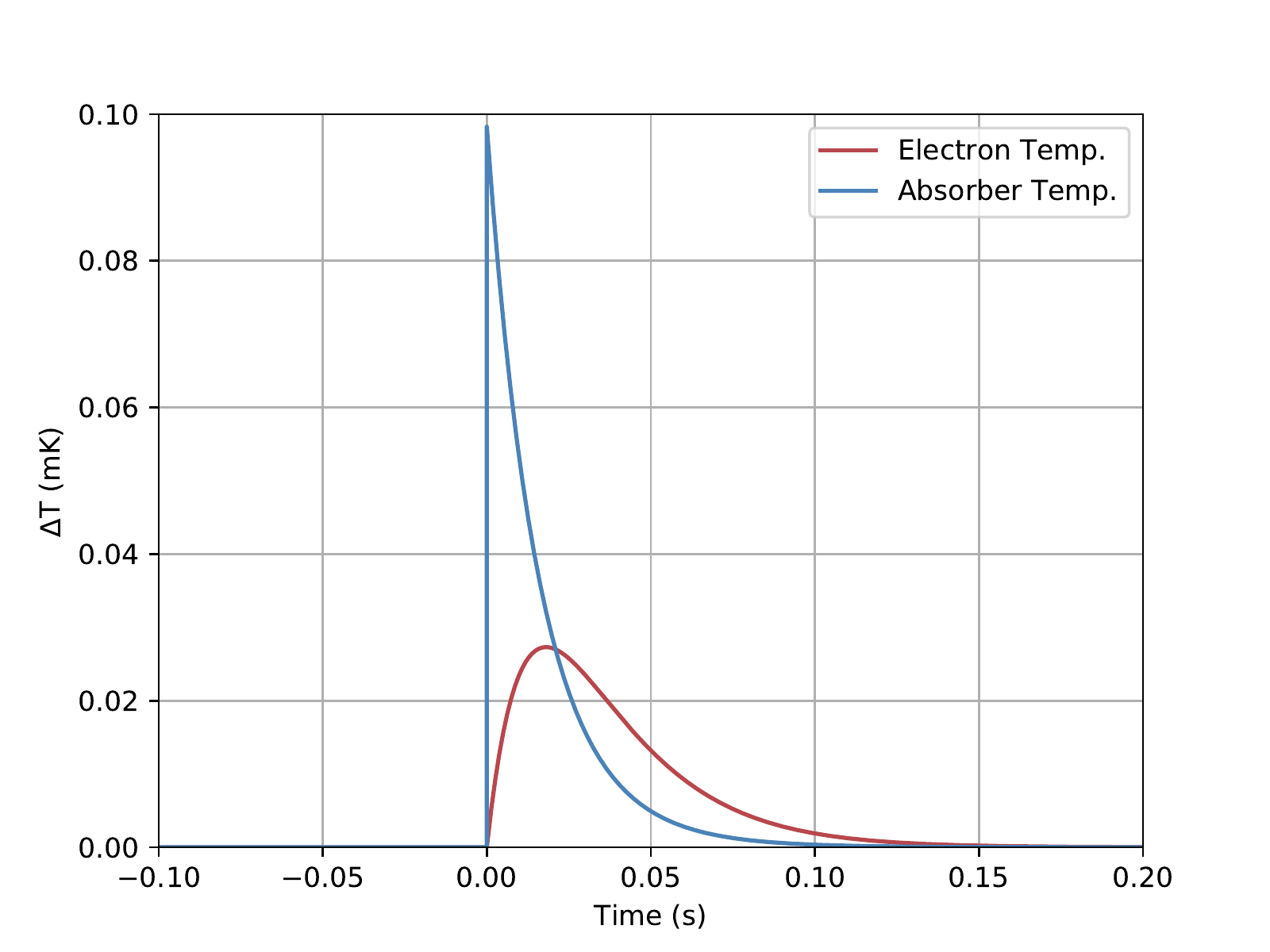}
    \caption{Simulated temperature variations in 1 cm$^3$ of solid xenon (absorber temperature) and germanium NTD thermistor (electron temperature) from a 1 MeV energy deposition with a bath temperature $T_{b}$ = 20 mK.}
    \label{pulse}
\end{figure}

With the determination of the boundary conductances, it is now possible to simulate the full bolometer model. For this, we use thermistor, bias circuit, and background power values from \cite{Pedretti}. For a 1 cm$^3$ sample of solid xenon, we approximate that the absorber is in contact with $\sim$1 cm$^2$ of substrate. Additionally, the heat capacity of the xenon is obtained from the Debye model. The results from a simulation of a 1 MeV deposition, with the heat bath temperature set to 20 mK, are shown in Figure \ref{pulse}. It should be noted that the decay time of this signal is about an order of magnitude lower than that of the thermal signals from the CUORE bolometers, but $\Delta T/T$ is of the same order of magnitude \cite{Pedretti, Vignati}. The quick decay of the thermal signal is mostly due to the large contact area between the absorber and the substrate/heat bath.

\section{Sample Growth and Characterization Setup}

Initial studies are underway at Drexel University to understand the growth of solid argon and xenon samples through a vapor deposition technique that might be adaptable to future in situ growth in a dilution refrigerator. The setup consists of an ARS DE-202 closed-cycle helium cryocooler, capable of reaching liquid helium temperatures, inside of a vacuum chamber. Gas is introduced into the system through a UHV leak valve. The gas will condense on a substrate that is thermally linked to the second stage of the cryocooler. It is estimated that a 1 cm$^3$ sample will take approximately one to two days to grow. Upon completion of growth studies, efforts will be focused on measuring the properties of charge transport and light yield. Though some of these parameters have been measured for xenon by \cite{Yoo:2015yza} at higher temperatures, little information is available at lower temperatures. The growth chamber is equipped with a fused quartz viewport which will be fitted with a photodetector module. While a UV-sensitive photodetector would suffice for xenon, this setup is capable of producing both xenon and argon samples. For this reason, a wavelength shifting material will be used to shift the vacuum ultraviolet light into the visible spectrum to be detected by conventional photodetectors. The results of these studies will inform design choices for the future noble solid bolometer experiments in the Drexel dilution refrigerator, which has a base temperature of 20 mK.

\section{Conclusions}

The microscopic anti-correlation between scintillation and ionization in noble liquid detectors is empirically well-modeled, but not understood from first principles. A solid xenon bolometer has the potential to shine light on this effect by measuring all three energy channels simultaneously. Little is known about noble solid elements as detector media, and current work at Drexel University is focused on studying the growth and characterization of these materials. Also, we are developing designs and models of solid argon and solid xenon bolometers. We have calculated conductance contributions from interfaces between materials to demonstrate the feasibility of reading out noble solid bolometers. The conductance values were calculated using the acoustic and diffuse mismatch models to give upper and lower bounds. These calculations were incorporated into a thermal simulation of a 1 cm$^3$ solid xenon sample with a germanium NTD thermistor at 20 mK, which showed that the temperature variations from MeV-scale energy depositions are reasonable for readout.

\end{document}